\documentstyle[12pt]{article}
\def\o{\over}

\def\bar{\overline}

\def\r{\gamma}
\def\a{\alpha}
\def\b{\beta}
\def\e{\epsilon}
\def\p{\pi}
\def\Im{{\rm Im}}
\def\Re{{\rm Re}}

\def\t{\tilde}
\def\bar{\overline}

\begin{document}
\setlength{\baselineskip}{8mm}
\begin{titlepage}
\begin{flushright}
UWThPh-1994-38\\
AUE-07-94\\
September 1994
\end{flushright}
\begin{center}
{\large\bf Neutron Electric Dipole Moment
in Two-Higgs-Doublet Model
\footnote{Talk given by M. Tanimoto at QCD94 in Montpellier, France on
 7-13 July 1994
and also predented by M. Matsuda at
the 27th International Conference
on High Energy Physics in Glasgow, Scotland on 20-27 July 1994.}}
{\bf T. Hayashi}$^\dagger$
       , {\bf Y. Koide}$^{\dagger\dagger}$\\
       ,  {\bf M. Matsuda}$^\sharp$
\footnote{E-mail:masa@auephyas.aichi-edu.ac.jp}
    and {\bf M. Tanimoto}$^\flat$
\footnote{Permanent address:Science Education Laboratory, Ehime University
, Matsuyama, 790 JAPAN}
\end{center}
\begin{center}
$^\dagger$Kogakkan University, Ise, Mie 516,
                     JAPAN\\
$^{\dagger\dagger}$Department of Physics, University of
            Shizuoka,  52-1 Yada, Shizuoka 422, JAPAN\\
$^\sharp$Department of Physics and Astronomy,
      Aichi  University of Education, Kariya 448, JAPAN\\
$^\flat$Institut f\"ur Theoretische Physik,
               Universit\"at Wien, A-1090 Wien, AUSTRIA
\end{center}

\begin{abstract}
The effect of the "chromo-electric" dipole moment  on
the electric dipole moment(EDM) of the neutron is studied in the
two-Higgs-doublet model.
   The Weinberg's operator $O_{3g}=GG\t G$
and the operator $O_{qg}=\bar q\sigma\t Gq$ are both investigated
in the cases of $\tan\b\gg 1$, $\tan\b\ll 1$ and $\tan\b\simeq 1$.
    The neutron EDM is considerably reduced due to the destructive
contribution with two light Higgs scalars exchanges.
\end{abstract}
\end{titlepage}
\section{Introduction}
  The electric dipole moment(EDM) of the neutron is of central importance
to probe a new origin  of $CP$ violation, because it is very small in SM
\cite{KM}($d_n^{SM} \simeq 10^{-30}-10^{-31}e\cdot cm$).
Begining with the papers of Weinberg
\cite{WB},
there has been considerably renewed
interest in the neutron EDM induced by $CP$ violation of the neutral
Higgs sector.
Some studies
\cite{GW,DG,BZ}
revealed the importance of the "chromo-electric" dipole
moment, which arises from the three-gluon operator
$GG\t G$ found by Weinberg
\cite{WB}
and the light quark operator $\bar q \sigma\t Gq$  introduced
by Gunion and Wyler
\cite{GW},
in the neutral Higgs sector.
Thus, it is important to study the effect of these operators systematically
in the model beyond  SM.
We study the contribution of above two operators to
the neutron EDM in the two-Higgs-doublet model(THDM)
\cite{GHKD}.
  The $3\times 3$ mass matrix of the neutral Higgs scalars is
carefully investigated in the typical three cases of $\tan\b\gg 1$,
$\tan\b\simeq 1$ and $\tan\b\ll 1$. In this model $CP$ symmetry is violated
through the mixing among $CP=+$ and $CP=-$ Higgs scalar states.
 \par
  In order to give reliable predictions
\cite{MG},
one needs the improvement on the
accuracy of the description of the strong-interaction hadronic  effects.
 Chemtob
\cite{C}
proposed a systematic approach
which gives the hadronic matrix elements of the higher-dimension operators
involving the gluon fields.
We employ his model to estimate the hadronic matrix elements of the  operators.

\section{$CP$ violation parameter in  THDM}

The simplest extension of SM is the one with the two Higgs  doublets
\cite{GHKD}.
This model      has the possibility
of the soft $CP$ violation in the neutral Higgs sector, which
does not contribute to the flavor changing neutral current in the $B$, $D$
and $K$ meson decays.
Weinberg
\cite{WB2}
has given the unitarity bounds for
the dimensionless parameters of the $CP$ nonconservation in THDM.
However, the numerically estimated
values of these parameters are not always  close to the Weinberg's
bounds
\cite{WB2}.
  Although it is difficult to estimate the magnitudes of the $CP$ violation
parameters ${\rm Im} Z_i(i=1,2)$ generally,
 we found that the neutral Higgs mass matrix is simplified in the extreme
cases of  $\tan\b\ll 1$, $\tan\b\simeq 1$ and $\tan\b\gg 1$, in which the
$CP$ violation parameters are easily calculated.
The $CP$ violation parameters ${\rm Im} Z_i^{(n)}$ are deduced to
\begin{eqnarray}
  {\rm Im} Z_1^{(k)}&=&-{\tan\b\o \cos\b}u_1^{(k)} u_3^{(k)} \ , \nonumber \\
  {\rm Im} Z_2^{(k)}&=& {\cot\b\o \sin\b}u_2^{(k)} u_3^{(k)} \ ,
 \end{eqnarray}
\noindent
where $u_i^{(k)}$ denotes the $i-$th component of the $k-$th
eigenvector of the $3 \times 3$ Higgs mass matrix and $\tan\b\equiv
v_2/v_1(v_{1(2)}$ is
the vacuum expectation value of $\Phi_{1(2)}^0$ giving the masses of
$d(u)$-quark sector).
\par
In this model, Higgs potential is generally given as
\begin{eqnarray}
V_H(\Phi_1,\Phi_2)
&=&{1 \o 2}g_1(\Phi_1^{\dag}\Phi_1-|v_1|^2)^2 \nonumber\\
&+&{1 \o 2}g_2(\Phi_2^{\dag}\Phi_2-|v_2|^2)^2 \nonumber\\
&+&g(\Phi_1^{\dag}\Phi_1-|v_1|^2)(\Phi_2^{\dag}\Phi_2-|v_2|^2) \nonumber\\
&+&g'|\Phi_1^{\dag}\Phi_2-v_1^*v_2|^2 \nonumber\\
&+&Re\{h(\Phi_1^{\dag}\Phi_2-v_1^*v_2)^2\} \nonumber\\
&+&\xi[{\Phi_1 \o v_1}-{\Phi_2 \o v_2}]^{\dag}[{\Phi_1 \o v_1}-
{\Phi_2 \o v_2}],
\end{eqnarray}
where the parameters satisfy the conditions
\cite{KAS}
\begin{eqnarray}
&g_1&\geq 0, \nonumber\\
&g_2&\geq 0, \nonumber\\
&g&> -\sqrt{g_1g_2}, \nonumber\\
&g&+g'-|h|\geq -\sqrt{g_1g_2}, \nonumber\\
&\xi&\geq 0, \nonumber\\
&g'&-|h|+{\bar \xi}\geq 0, \nonumber\\
&{\bar \xi}&-g\geq -\sqrt{g_1g_2} \quad ({\rm where}
\ {\bar \xi}\equiv {\xi \o |v_1v_2|^2}).
\end{eqnarray}
It is noted that, in the case of MSSM, SUSY imposes the conditions on
the parameters
\begin{eqnarray}
&g_1=g_2={1 \o 4}({g_W}^2+{g'_W}^2), \nonumber\\
&g={1 \o 4}({g_W}^2-{g'_W}^2), \nonumber\\
&g'=-{1 \o 2}{g_W}^2, \nonumber\\
&h=0.
\end{eqnarray}
Here $h$=0 means that in MSSM $CP$ violation is not caused throygh Higgs
sector. The simplest SUSY extention from MSSM
that can have CP violation in the
Higgs sector is also discussed
\cite{EG}.
\par
 Let us estimate $u_i^{(k)}$ by studying the Higgs mass matrix ${\bf M^2}$
whose components are
\begin{eqnarray}
M_{11}^2&=&2g_1|v_1|^2+g'|v_2|^2+{\xi+\Re(hv_1^{*2}v_2^2)\o|v_1|^2}\ ,
\nonumber\\
M_{22}^2&=&2g_2|v_2|^2+g'|v_1|^2+{\xi+\Re(hv_1^{*2}v_2^2)\o|v_2|^2}\ ,
\nonumber\\
M_{33}^2&=&(|v_1|^2+|v_2|^2) \left [g'+
              {\xi-\Re(hv_1^{*2}v_2^2)\o|v_1v_2|^2}\right ]\ ,\nonumber\\
M_{12}^2&=&|v_1v_2|(2g+g')+{\Re(hv_1^{*2}v_2^2)-\xi\o|v_1v_2|}\ , \nonumber \\
M_{13}^2&=&-{\sqrt{|v_1|^2+|v_2|^2}\o|v_1^2v_2|}\Im(hv_1^{*2}v_2^2)\ , \\
M_{23}^2&=&-{\sqrt{|v_1|^2+|v_2|^2}\o|v_1v_2^2|}\Im(hv_1^{*2}v_2^2)\nonumber
\ .
\end{eqnarray}
\noindent
As a phase convension, we take $h$ to be real and
\begin{equation}
  v_1^{*2} v_2^2=|v_1|^2|v_2|^2\exp(2i\phi) \ .
\end{equation}
  At first, we consider the case of $\tan\b\gg 1$ with
 retaining the order of $\cos\b$ and  setting $\cos^2\b=0$
and  $\sin\b=1$. Then,
the mass matrix becomes simple, so
 the eigenvectors of ${\bf M^2}$ in Eq.(5) are easily obtained
as follows:
\begin{eqnarray}
u^{(1)}&=&\{\matrix{\cos\b-\e\sin\b, &
             -\sin\b, & 0 }\} ,\\
u^{(2)}&=&\{\matrix{\sin\b c_\phi, & (\cos\b-\e\sin\b)c_\phi,
              & -s_\phi }\} , \nonumber\\
u^{(3)}&=&\{\matrix{\sin\b s_\phi, &(\cos\b -\e\sin\b) s_\phi,
              & c_\phi }\} , \nonumber
\end{eqnarray}
\noindent where $c(s)_\phi\equiv \cos(\sin)\phi$ and
\begin{equation}
 \e \simeq {2(\bar\xi-g-g_2)\o \bar\xi+g'-2g_2}\cos\b \ .
\end{equation}
\noindent
The diagonal masses are given as
\begin{equation}
M_1^2=2g_2,\  M_2^2=g'+\bar\xi+h,
       \  M_3^2=g'+\bar\xi-h
\end{equation}
\noindent               in the $v^2\equiv v_1^2+v_2^2$ unit.
The lightest Higgs scalar to yield  $CP$ violation is the  second Higgs
  scalar with the mass $M_2$ since $\bar\xi$ is positive from Eq.(3)
and we take $h$ to be negative as convention.
 The Higgs scalar with $M_1$ does not contribute to  $CP$ violation
because of $u_3^{(1)}=0$. The absolute values of  $g'$
is expected to be $O(1)$, but $h$ seems to be small as estimated
in  some works
\cite{C2,L}.
For example Froggatt et al. give the numerical values for the parameters
in the case of $\tan\b \gg 1$ by using infrared fixed point  analysis through
the renormalization group equations as
\begin{eqnarray}
g_1\simeq 0.96,\ g_2\simeq 0.88, \ g\simeq 0.82 \nonumber\\
g'\simeq -1.20,\ h\simeq -0.09 .
\end{eqnarray}
 Therefore, the masses $M_2$ and $M_3$ may be almost degenerated. Then,
$CP$ violation is reduced by the cancellation between the two different
  Higgs  exchange contributions $\Im Z_i^{(2)}$ and $\Im Z_i^{(3)}$
since $u^{(2)}_i u^{(2)}_3$ and
  $u^{(3)}_i u^{(3)}_3$(i=1,2) have same magnitudes with opposite signs.
 Thus, it is noted that
      the lightest single Higgs exchange approximation gives miss-leading
of $CP$ violation in the case of $\tan\b\gg 1$.\par
 For  ${\Im}Z_1$, our result reaches the Weinberg bound, but for  ${\Im}Z_2$
 the our calculated value is suppressed compared with
 the Weinberg bound in the order of  $1/\tan\b$. \par
$CP$ violation in the case of $\tan\b\ll 1$ is similar to the one of $\tan\b\gg
   1$.
For  ${\Im}Z_2$, our numerical result reaches the Weinberg bound, while for
${\Im}Z_1$
 the calculated value is suppressed from the Weinberg bound in the order of
 $\tan\b$. The relative sign between ${\Im}Z_1$ and ${\Im}Z_2$ is just
 the  same as  in the case of $\tan\b \gg 1$.\par
The last case to be considered is   of $\tan\b\simeq 1$.
In this mass matrix, the off diagonal components are very small compared to the
diagonal ones because $g_1\simeq g_2$ is suggested by some
 analyses
\cite{C2,L}
and $h$ is also small as in the case of $\tan\b \gg 1$.
 We can calculate  ${\Im}Z_i$ by fixing both values of  $h$ and $M_2/M_3$.
For  both ${\Im}Z_2$ and   ${\Im}Z_1$,
 the calculating values are roughly 1/3 of the Weinberg bounds.
The relative sign between ${\Im}Z_1$ and ${\Im}Z_2$
is opposite.\par
\section{Formulation of the neutron EDM}
The low energy $CP$-violating interaction is described by an effective
Lagrangian,
\begin{equation}
L_{CP}=\sum_i C_i(M,\mu)O_i(\mu) \ ,
\end{equation}
\noindent where $O_i$ are
 the three gluon operator with the dimension six
and the quark-gluon operator with the dimension five
as follows:
  \begin{eqnarray}
  O_{qg}(x)&=&-{g_s^3\o 2}\bar q\sigma_{\mu\nu}\tilde G^{\mu\nu} q \ ,
\nonumber \\
  O_{3g}(x)&=&-{g_s^3\o 3}f^{abc}\tilde G^a_{\mu\nu}G^b_{\mu\a}G^c_{\nu\a} \ ,
\end{eqnarray}
\noindent where $q$ denotes $u,d$ or $s$ quark.  The QCD corrected
coefficients $C_i$ are given by the two-loop calculations in Refs.
\cite{WB,GW}.
The coefficients $C_i$ are given as
\begin{eqnarray}
C_{ug}&=&{\sqrt{2}G_Fm_u \o 64\pi^4}\{f({m_t^2 \o m_H^2})
+g({m_t^2 \o m_H^2})\}\Im Z_2({g_s(\mu) \o g_s(M)})^{-{74 \o 23}},\nonumber\\
C_{dg}&=&{\sqrt{2}G_Fm_d \o 64\pi^4}\{f({m_t^2 \o m_H^2})\tan^2\b \Im Z_2
\nonumber\\
&-&g({m_t^2 \o m_H^2})\cot^2\b \Im Z_1\}({g_s(\mu) \o g_s(M)})^{-{74 \o 23}},
\nonumber\\
C_{3g}&=&{\sqrt{2}G_F \o (4\pi)^4}\Im Z_2h({m_t^2 \o m_H^2})
({g_s(\mu) \o g_s(M)})^{-{108 \o 23}},
\end{eqnarray}
where the functions $f(x),g(x),h(x)$ are deduced from loop integral
as given in Refs.
\cite{WB,GW}.
\par
 For the strong interaction hadronic effect, the systematic technique has
been developed  by Chemtob
\cite{C}
in the operator with the higher-dimension involving the gluon fields.
The hadronic matrix elements of the two operators
 are approximated by the intermediate states with the single nucleon pole
 and the  nucleon plus one pion. Then, the nucleon matrix elements are
 defined as
\begin{eqnarray}
  \langle N(P)|O_i(0)|N(P)\rangle = A_i\bar U(P)i\r_5 U(P), \nonumber \\
  \langle N(P')|O_i|N(P)\p(k)\rangle= B_i\bar U(P')\tau^a U(P) \ ,
\end{eqnarray}
\noindent
where $U(P)$ is the normalized nucleon Dirac spinors
               with the four momuntum $P$.
Using $A_i$ and $B_i(i=ug,dg,sg,3g)$, the neutron EDM, $d_n^\r$,
are written as
\begin{equation}
  d_n^\r={e\mu_n\o 2 m_n^2}\sum_i C_i A_i +
  F(g_{\p NN})\sum_i C_i B_i    \ ,
\end{equation}
\noindent where $\mu_n$ is the neutron anomalous magnetic moment.
 The $F(g_{\p NN})$ was given by calculating the pion
and nucleon loop corrections using the chiral Lagrangian
for $N\p\r$
\cite{C}.
The coefficients $A_i$ and $B_i$ were given by the large $N_c$
current algebra and the $\eta_0$ meson dominance
\cite{C}.
\section{Numerical results of the neutron EDM}
 Let us begin with discussing the numerical results in
 the case of $\tan\b\gg 1$.
  The contributions of $O_{ug}$ and $O_{3g}$ are
 are negligibly small because the $CP$ violation parameters are roughly
estimated as
\begin{eqnarray}
\Im Z_2^{(2)}\simeq-\Im Z_2^{(3)}\simeq{1 \o \tan^2\b} \ll \Im Z_1^{(2,3)},
\nonumber\\
\Im Z_1^{(3)}\simeq -\Im Z_1^{(3)}\simeq {1 \o 2}\tan^2\b.
\end{eqnarray}
   The main contribution
follows from the one of $O_{dg}+O_{sg}$, in which the operator $O_{sg}$
is dominant due to the $s$-quark mass.
The coefficient $C_{sg}$ is
\begin{eqnarray}
C_{sg}&=(const.)\times m_s\{f({m_t^2 \o m_{H_2}^2})
-f({m_t^2 \o m_{H_3}^2}) \nonumber\\
&-{1 \o 2}g({m_t^2 \o m_{H_2}^2})+{1 \o 2}g({m_t^2 \o m_{H_3}^2})\}.
\end{eqnarray}
  As the mass difference of these
two Higgs scalar masses becomes smaller, the neutron EDM  is considerably
reduced since the second Higgs scalar exchange  contributes
in the opposite sign to the  lightest Higgs scalar one as shown in the
above equation.
 Thus, it is found that the second lightest
Higgs scalar also significantly contributes to  $CP$ violation.\par
In the case of  $\tan\b\ll 1$,
the contributions of $O_{ug}$ and $O_{3g}$ become very large due to  the
large $\Im Z_2$.
 However, these contribute to the neutron EDM  in opposite signs,
so they almost cancel each other.
The remaining contribution is the one of $O_{dg}+O_{sg}$.\par
In the case of $\tan\b\simeq 1$,
the dominant contribution
 is the one of $O_{dg}+O_{sg}$.
In both regions of the large and small $m_{H2}/m_{H3}$,
the predicted neutron EDM is reduced.
At  $m_{H2}/m_{H3}\simeq 1$, the cancellation mechanism by the
second lightest Higgs scalar operates well, while
  around $m_{H2}/m_{H3}\simeq 0$,
the large mass difference of the two Higgs scalars leads to the small mixing
between the scalar and pseudscalar Higgs bosons.
\section{Summary}
We have studied  the effects  of the
 four operators $O_{ug}$,  $O_{dg}+O_{sg}$ and $O_{3g}$ on the neutron EDM.
The contribution of $O_{sg}$  dominates over that of other operators. Moreover,
the contributions of $O_{ug}$ and $O_{3g}$ cancel out each other
due to their opposite signs.  This qualitative situation does not
depend on the detail of the strong interaction hadronic model.
Thus, the Weinberg's three gluon operator
is not a main source of the neutron EDM in THDM.
The CP violation mainly follows from the two light neutral Higgs  scalar
exchanges. Since these two exchange contributions are of
opposite signs, the $CP$ violation is considerably reduced
 if the mass difference of the two Higgs scalars is small.
 Since our predicted neutron EDM lies around the present
 experimental bound,
  its experimental improvement reveal the new physics
beyond SM.
The present upper limit for $d_n^\gamma$ is $8 \times 10^{-26}e\cdot cm$ which
was given at the 26th ICHEP. Historically to reduce one order of
magnitude for upper limit experimentally, it has taken almost 10 years.
We hope that the rapid experimental reduction of upper limit will be performed
and that the finite value will be reported at the close ICHEP.


\begin{thebibliography}{20}
\bibitem{KM}  M.\ Kobayashi and T.\ Maskawa, Prog. Theor.
  Phys. {\bf 49}(1973) 652.
\bibitem{WB} S.\ Weinberg, Phys. Rev. Lett. {\bf 63}(1989) 2333.
\bibitem{GW} J.F.\ Gunion and D.\ Wyler, Phys.Letts.{\bf 248B} (1990)170.
\bibitem{DG} A.\ De R\'ujula, M.B.\ Gavela, O.\ P\`ene and F.J.\ Vegas,
Phys.Lett.{\bf 245B} (1990) 640;
    N-P.\ Chang and D-X.\ Li, Phys. Rev. {\bf D42}(1990)871;
D.\ Chang, T.W.\ Kephart, W-Y.\ Keung and T.C.\ Yuan,
Phys. Rev. Lett. {\bf 68}(1992)439.
\bibitem{BZ} S.M.Barr and A.Zee, Phys. Rev. Lett. {\bf 65}(1990)21;
S.M.\ Barr, Phys. Rev. Lett. {\bf 68}(1992)1822, Phys. Rev. {\bf D47}(1993)
2025.
\bibitem{GHKD} J.F.\ Gunion, H.E.\ Haber, G.L.\ Kane
    and S.\ Dawson,  {\it "Higgs Hunter's Guide"},  Addison-Wesley, Reading,
MA(1989).
\bibitem{MG} A.\ Manohar and H.\ Georgi, Nucl.Phys.{\bf B234} (1984)189.
\bibitem{C} M.\ Chemtob, Phys. Rev. {\bf D45}(1992)1649.
\bibitem{WB2} S.\ Weinberg, Phys. Rev. {\bf D42}(1990)860.
\bibitem{KAS} B.\ Kastening, Private communications and also see the preprint(
hep-ph@9307225).
\bibitem{EG} J.\ Ellis, J.F.\ Gunion, H.E.\ Haber, L.\ Roszkowski and
F.\ Zwirner, Phys. Rev. {\bf D39}(1989)844; See also Refs.\cite{BZ}.
\bibitem{C2} M.\ Chemtob, Z.Phys. {\bf C60}(1993)443.
\bibitem{L} M.A.\ Luty, Phys. Rev. {\bf D41}(1990)2893;
  C.D.\ Froggatt, I.G.\ Knowles and R.G.\ Moorhouse,
 Phys. Lett.   {\bf 249B} (1990)273.
\end{thebibliography}
\end{document}